\documentclass[twocolumn]{aastex62}

\usepackage{amsmath}
\usepackage{color}
\usepackage{graphicx}
\usepackage{hyperref}
\hypersetup{pdfauthor=Judit Szul\'agyi}
\hypersetup{backref=true, pagebackref=true, hyperindex=true, breaklinks=true,colorlinks=true,urlcolor=blue, linkcolor=blue,  citecolor=blue,pagecolor=red, bookmarks=true, bookmarksopen=true}

\received{August 10, 2018}
\revised{XX, 2018}
\accepted{XX, 2018}
\submitjournal{ApJL}

\shorttitle{Satellite-formation around Ice Giants}
\shortauthors{Szul\'agyi et al.}

\begin{document}

\title{IN SITU FORMATION OF ICY MOONS OF URANUS AND NEPTUNE}

\correspondingauthor{Judit Szul\'agyi}
\email{judit.szulagyi@uzh.ch}

\author[0000-0001-8442-4043]{Judit Szul\'agyi}
\affil{Center for Theoretical Astrophysics and Cosmology,\\
Institute for Computational Science, University of Zurich,\\
Winterthurerstrasse 190, CH-8057 Zurich, Switzerland.}
\affil{Institute for Particle Physics and Astrophysics, \\
ETH Zurich, Wolfgang-Pauli-Strasse 27, 8093, Zurich, Switzerland}

\author{Marco Cilibrasi}
\affil{Center for Theoretical Astrophysics and Cosmology,\\
Institute for Computational Science, University of Zurich,\\
Winterthurerstrasse 190, CH-8057 Zurich, Switzerland.}

\author{Lucio Mayer}
\affil{Center for Theoretical Astrophysics and Cosmology,\\
Institute for Computational Science, University of Zurich,\\
Winterthurerstrasse 190, CH-8057 Zurich, Switzerland.}



\begin{abstract}
Satellites of giant planets thought to form in gaseous circumplanetary disks (CPDs) during the late planet-formation phase, but it was unknown so far whether smaller mass planets, such as the ice giants could form such disks, thus moons there. We combined radiative hydrodynamical simulations with satellite population synthesis to investigate the question in the case of Uranus and Neptune. For both ice giants we found that a gaseous CPD is created at the end of their formation. The population synthesis confirmed that Uranian-like, icy, prograde satellite-system could form in these CPDs within a couple of $10^5$ years. This means that Neptune could have a Uranian-like moon-system originally that was wiped away by the capture of Triton. Furthermore, the current moons of Uranus can be reproduced by our model without the need for planet-planet impact to create a debris disk for the moons to grow. These results highlight that even ice giants -- that among the most common mass-category of exoplanets -- can also form satellites, opening a way to a potentially much larger population of exomoons than previously thought.
\end{abstract}



\section{Introduction} \label{sec:intro}

In our Solar System satellites can be found mainly around gas giant planets. As a scaled-down version of planet formation, moons are also assembled within disks: in the gaseous circumplanetary disks (CPD) surrounding the giant planets during their last stage of formation. The larger moons around Jupiter and Saturn are thought to form in gaseous CPDs \citep{LS82,PB89,CW09}. In contrast, terrestrial planets, such as the Earth and Venus have too low mass to gather such a disk in the first place, they can only form an envelope which might turn into a primitive atmosphere \citep{Ormel15}. The reason why Earth still has a moon is believed to be a result of a planet-planet impact, after which the ejected material formed a debris disk around our planet, where eventually the Moon assembled \citep{HD75,CW76}. The planetary mass threshold below which gaseous CPD formation -- hence satellite-formation -- cannot occur is still unknown \citep{AB09}. 

In the past years radiative hydrodynamic simulations of CPDs showed that whether a CPD or an envelope forms is not only a question of planetary mass, but also the temperature \citep{Szulagyi16,Szulagyi17gap,Sz17b}. These simulations revealed that the cooler is the planet (or the surrounding accreting flow), the more likely it is that a disk can form. Satellite formation thus depends on both the planetary mass and the temperature. As planets radiate away their formation heat, they constantly and rapidly cool within the first few million years of their lifetime, therefore whether and when they form a disk is a question of age (and thermal history) as well. Lying between the gas giant and terrestrial planet regime, it was so far unknown whether ice giants, like Uranus and Neptune could ever form a gaseous CPD and therefore capable of forming their satellites there. 

Uranus has five major, regular, prograde moons that have nearly circular orbits and low inclinations, which suggests that they formed in a disk \citep{ME03a}. Another model suggested an expanding, tidal disk made of solids \citep{CCh12}. However, because the planet has an obliquity of 98 degrees, it was assumed that there has been an impact with an Earth-sized object \citep{Safronov66,HW82,Slattery92}. This impact could have resulted a debris disk around the planet (similar to the case of Earth) where its moons formed \citep{Dermott84,Stevenson84,Mousis04}. The debris disk of the giant impact however had to be retrograde \citep{Morbidelli12}, and the strong impact would evaporate the ice from the ejected debris \citep{Mousis04} hence the resulted satellites would be poor in water ice in contrast the observations. The Uranian satellites are in fact made of $\sim$50\% water ice and $\sim$50\% rock. Instead of the major impact scenario, it was suggested that perhaps multiple, smaller impacts caused the tilt and formed a debris disk around the planet \citep{Morbidelli12}, however, to still form prograde satellites, this model require fine tuning in order to work. Another possibility to tilt Uranus is via secular resonances by a massive moon over a long period of time \citep{BL10}.

In case of Neptune, there is only one major moon, Triton, that has 99\% of the mass of the entire satellite system. Based on its composition, 157 degree inclination and retrograde orbit, Triton is almost certainly a captured Kuiper-belt object \citep{McKinnon95}. The capture could have dynamically distorted the original satellite system of Neptune, if there was any \citep{Rufu17}. 

So far studying satellite-formation around Uranus and Neptune was particularly challenging. To address CPD formation realistically, the simulations must have the following two characteristics: (1) sufficiently high resolution to resolve the Hill-sphere, because under-sampling will lead to incorrect physics that will result in envelope formation, and (2) treatment of temperature e.g. with radiation-hydrodynamics, since the temperature in the planet vicinity will also control whether or not a disk can form \citep{Szulagyi16,Szulagyi17gap,Cimerman17}. So far there was one hydrodynamics work trying to address the CPD formation around Neptune-sized planets, but \citet{Wang14hydro} did not have any temperature treatment. Particle-based simulations could only address the debris disk hypothesis so far, this was done by a number of works modeling a potential planet-planet impact \citep{Slattery92,Mousis04,Korycansky90}. 

In this work, we go beyond these previous studies using radiation hydrodynamic simulations with unprecedented resolution in the vicinity of the ice giants. We investigate whether a gaseous CPD could have formed originally around Uranus and Neptune. As a next step, we checked whether satellite-formation could have occurred in those disks with the use of satellite population synthesis.  

\section{Methods} \label{sec:methods}

\subsection{Hydrodynamical Simulations}

We ran three-three hydrodynamic simulations for Uranus and Neptune with the JUPITER code \citep{Szulagyi16} developed by F. Masset \& J. Szul\'agyi. The code is three-dimensional, grid-based, solving the Euler equations together with radiative dissipation via the flux the limited diffusion approximation method \citep{Kley89,Com11}. In each case there is a spherical coordinate system with a star in the center, surrounded by a gas circumstellar disk. The planets were treated as point masses at the location of the current Uranus (19.2 AU) and Neptune (30.1 AU). The circumstellar disk radius ranged between 7.7 AU and 45.8 AU for Uranus, and 12.0 AU and 71.8 AU for Neptune. Radially 215 cells, azimuthally 618 cells were used on the base mesh. The initial disk opening angle was 5.68 degrees (from the midplane to the disk surface, using 20 cells), but the disk got thinner after reaching thermal equilibrium with heating and cooling effects. The initial surface density was $16.29 \,\rm{g/cm^2}$ at the location of Uranus and $6.63 \,\rm{g/cm^2}$ at Neptune’s, considering flat disks. This setup is corresponding to roughly half of the Minimum Mass Solar Nebula \citep{Hayashi} densities, mimicking the very last stage of planet formation within the Solar System when the satellites were formed. Given that real circumstellar disks are dissipating in a nearly exponential fashion at the end of their lifetime, our choice on setting half of the Minimum Mass Solar Nebula was just an educational guess for a mean value that is definitely a mass what the Solar Nebula eventually had at the late stage of its lifetime. 

The equation of state in these simulations were of an ideal gas: $p=(\gamma-1)E_{int}$  connecting the (p) pressure with the internal energy ($E_{int}$) via the adiabatic exponent ($\gamma=1.43$). The code solves the viscous stress tensor for a constant, kinematic viscosity, that equals to $1.95 \times 10^{14} \,\rm{cm^2/s}$ in the case of Uranus, and $2.44\times 10^{14} \,\rm{cm^2/s}$ for Neptune. These values are corresponding to a reasonable, low alpha-like viscosity that scales with the semi-major axis from the star, which includes that planets form in dead-zones. In the simulations the gas can heat up through viscous heating, adiabatic compression and cool through radiative dissipation and adiabatic expansion. The Rosseland-mean-opacities used in the radiation-hydrodynamics simulations were constructed self-consistently from frequency dependent dust opacities computed with a version of the Mie code from \citet{BH84}. The dust consisted of 40\% water, 40\% silicates and 20\% carbonaceous material \citep{Zubko96,Draine03,Warren84}, assuming spherical, compact, micron sized grains. The star was assumed to have solar properties and the dust-to-gas ratio was chosen to be 1\%. The opacity table contains the evaporation of the different dust components: 170\,K for water, 1500\,K for silicates, and 2000\,K for carbon, respectively. Above 2000K the gas opacities were used from \citet{BL94}. This method ensures that even though the dust is not treated explicitly in the gas hydrodynamic simulations, its effect on the temperature is taken into account through the dust opacities. The mean-molecular-weight was set to 2.3 to be consistent with the solar composition. 

To have sufficient resolution in the circumplanetary disks, we increased the resolution in this area with nested meshes. We placed seven additional grids around the ice-giants, reaching a final resolution of $1.38 \times 10^{-3}$ AU (8.1 $\rm{R_p}$) for Uranus and $2.17 \times 10^{-3}$ AU (13.1 $\rm{R_p}$) for Neptune. For the planets, we fixed their mass, radius and temperature similarly to \citet{Szulagyi17gap}. The 8 cells around the point-masses were the extent of the planet, where the temperatures were fixed to 1000\,K, 500\,K and 100\,K in the three different simulations for Uranus and for Neptune, corresponding to the late surface temperatures of these planets during their formation around 3-4 million years \citep{FN10,Nettelmann16}. This temperature sequence was set in order to mimic how the planet radiates away its formation heat, up until it is fully formed. This is only true in the very late stage of circumplanetary disk evolution, close to the time when the circumstellar disk has dissipated away. 

\subsection{Population Synthesis}

To examine satellite formation within the CPDs of the hydrodynamical simulations, we used population synthesis. The technique was already successfully used in our previous work \citet{Cilibrasi18} for moon-formation around Jupiter, hence we only summarize here the main points of our method.

We azimuthally averaged the density and temperature filed on the midplane from the hydrodynamic simulations for the coldest planet case (100K) when a circumplanetary disk formed. Satellites can only form in disks, not in envelopes, therefore we only considered the disk cases. In the semi-analytical model the CPDs have the same properties as in the hydrodynamic simulation: same viscosity, adiabatic index, and mean-molecular-weight. The angular velocity of the gas, the scale-height of the disk, and the sound-speed was then computed from the local temperature and density values of the disk and using the common one-dimensional model for disks \citep{Pringle81}. Due to the fact that the hydrodynamic simulation only computes the gas distribution, we assumed that the dust distribution is the same, but only a fraction of it in mass controlled by the dust-to-gas ratio parameter. The temperature of the dust was assumed to be the same as the gas temperature, assuming perfect thermal equilibrium. The CPD ranged between 1 $\rm{R_p}$ and 500 $\rm{R_p}$, according to the hydrodynamical simulation's CPD radius. The population synthesis includes the CPD evolution (that it cools through radiative dissipation and changes is mass), the continuous mass infall from the circumstellar disk \citep{Szulagyi14} as well as the mass loss: accreting to satellites and to the planet. The net mass infall was also computed from the hydrodynamic simulations: $8.4 \times 10^{-8} \,\rm{M_{Sun}/yr}$ for Uranus, and $7.4 \times 10^{-8} \,\rm{M_{Sun}/yr}$ for Neptune. These infall rates then decreased exponentially, as the circumstellar disk also dissipates exponentially in the last phase of disk evolution and the feeding to the CPD changes accordingly. 

We ran 25000 individual calculations where in each case varied 4 parameters: 
\begin{enumerate}
\item Disk dispersion timescale: different circumstellar disks dissipate on different timescales based on observations, hence we chose to vary this parameter between 0.1 and 1 million years, which is roughly 1/10th of the total lifetime of the circumstellar disk and hence probably of the circumplanetary disk as well.
\item Dust-to-gas ratio: it ranges on a wide range in circumstellar disks \citep{Ansdell16} and according to dust coagulation studies the CPD can be quite dust rich \citep{DSz18}. We therefore varied this parameter between 1\% and 50\%. 
\item Dust refilling timescale: how quickly the dust is accreted and depleted within the CPD, how quickly it reaches again its equilibrium profile. This parameter was varied between 100 years and 1 million years \citep{Cilibrasi18}. 
\item Distance from the planet where the first seeds of satellites are forming. It was varied randomly between 50-150 planetary-radius.
\end{enumerate}

At the beginning of simulation, the algorithm creates a new protosatellite with a mass of $10^{-7} \,\rm{M_p}$, which is a seed size that quickly forms from the incoming micron sized dust, via coagulation \citep{DSz18}. Then it accretes mass from the dust disk with a common analytical prescription \citep{Greenberg91}. This accreted mass will be then subtracted from the dust disk density. The protosatellites that form will feel the torque exerted on them by the gas disk and start to migrate. As long as they are small and do not open a gap in the gas distribution of the CPD, they migrate according to the type I regime, computed with the Paardekooper-formula \citep{Paardekooper1, Paardekooper2}. In rare cases when they grow to such large mass that can open a gas-gap, they then enter the type II regime of migration and change their orbit on the viscous timescale \citep{Duffell15}. Sometimes the migration is so fast that the protosatellites will be accreted into the planet. In these cases, we kept track of the lost satellite-mass, in order to know what amount of solids the ice giants were ``polluted". The population synthesis simulation stopped when the CPD dissipated, because then the migration of the satellites stopped the and the moon-system that formed was considered final. 

\section{Results} \label{sec:results}

\subsection{Uranus}

\begin{figure}
\includegraphics[width=\columnwidth]{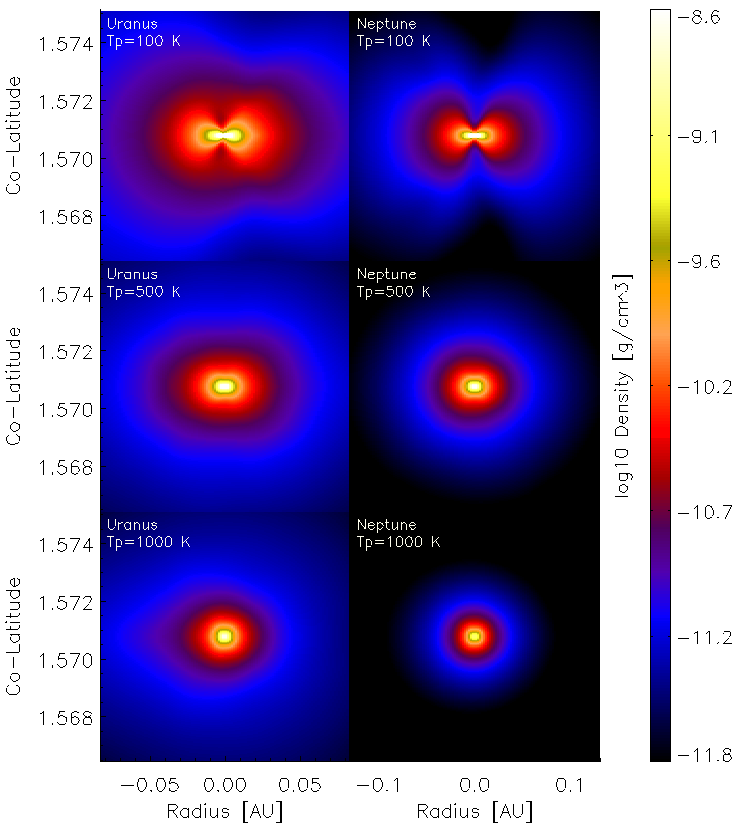}
\caption{Zoom to the circumplanetary disk around Uranus (left column) and Neptune (right column). The different rows of the density maps correspond to various planetary surface temperatures mimicking the late stages of ice giant formation when these planets were rapidly cooling.}
\label{fig:hydro}
\end{figure}

Owing to the high resolution we could achieve in the planet vicinity within the hydrodynamic simulations, we were able to determine that disks are forming when the planet temperature dropped below 500 K (Figure \ref{fig:hydro}). When steady state of the simulation was reached, the mass of the CPD was only $7.2\times 10^{-4} \,\rm{M_{Uranus}}$, measured based on the rotational profile and a corresponding isodensity surface. This is only slightly larger than the integrated mass of the moons ($\sim 10^{-4} \,\rm{M_{Uranus}}$). However, as it was described earlier, the CPD is not a closed reservoir of mass, it is continuously fed by the circumstellar disk with gas and micron-sized dust \citep{Szulagyi14,FC16}. Some of this mass is accreted to the planet and to the moons, some is flowing back to the circumstellar disk, maintaining a sort of equilibrium with the circumstellar disk mass \citep{Szulagyi17gap}. We measured the net influx rate to the circumplanetary disks from the hydrodynamic simulations by defining a box around the CPD and calculating the sum of the inflowing and outflowing gas, finding $8.4\times 10^{-8} \,\rm{M_{Sun}/yr}$, therefore over the years, there is enough mass flowing through the CPD to eventually build the current satellites of Uranus. 

\begin{figure*}
\includegraphics[width=18cm]{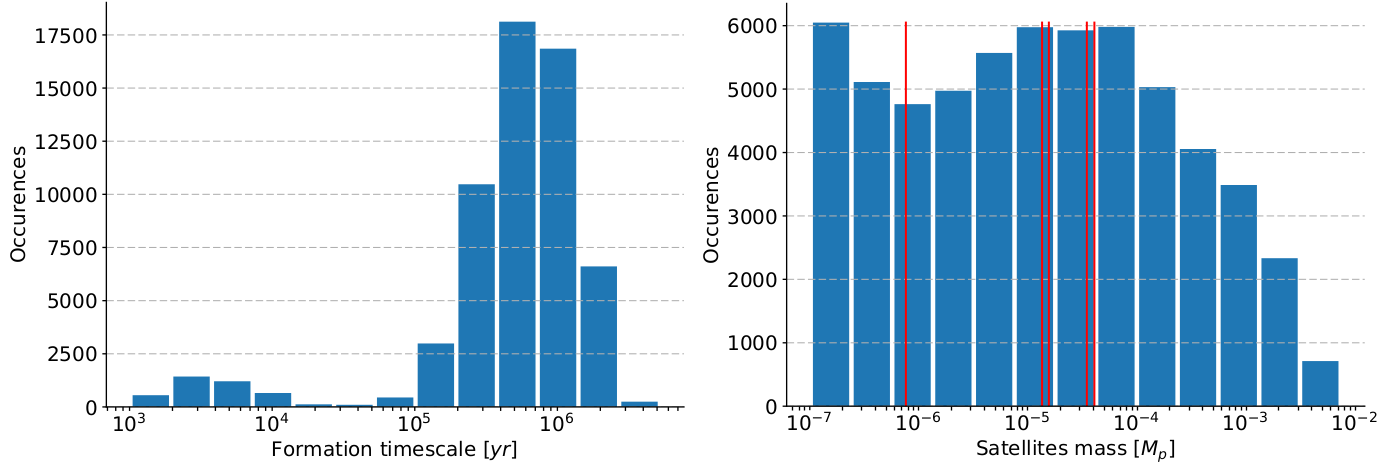}
\caption{Formation timescales of the moons around Uranus in our population synthesis (left panel). Mass distribution of the formed satellite population around Uranus, where the red lines represent its current, major five moons (right panel).}
\label{fig:uranus}
\end{figure*}

We tested with satellite population synthesis whether the CPD with planetary surface temperature of 100\,K could form moons. We found that in most cases of the population, there is a satellite-system forming. Most of the moons formed over only $\sim$500000 years (Fig. \ref{fig:uranus} left panel). While some of the satellites migrated inward in the disk up until they got accreted into the planet, the migration velocities were slow enough, that only 25\% of the population was lost. Often sequential satellite formation happened: after the lost satellites still new ones could form until the CPD has dissipated. All the satellites were formed at a location of the disk where the temperature was below water freezing point, hence the building blocks of these moons can contain significant fraction of ices to make icy moons, such as the current ones around Uranus. The masses of the formed satellites spread across several orders of magnitude, mainly in the smaller mass regime between $10^{-7}$ and $10^{-4} \,\rm{M_{Uranus}}$ (Fig. \ref{fig:uranus} right panel). On Figure \ref{fig:uranus} we also mark with red stripes the current masses of the major five satellites. Clearly, the moon-system of Uranus as we see today can be reproduced by our model. In 5.8\% of the cases we got out systems with 4 or 5 satellites and with a total mass between 0.5 and 2 times the current Uranian satellites. Considering the distances from the central planet, we found that in about 18\% of the cases they are comparable to the orbital radii of the current five Uranian satellites. Both conditions together occur in 5.1\% of all the cases.

We investigated the impact of different physical ingredients in the model by varying only one parameter at a time. We found that, the masses of the final moons basically linearly scales with the dust-to-gas ratio, hence in order to reproduce the major five satellites, a dust-to-gas ratio of 7\% or higher is needed within the CPD. Moreover, the dust refilling timescale has to be 10000 years, or shorter, meaning relatively quick dust evolution, which can easily happen according to a recent study of dust-coagulation in CPDs \citep{DSz18}.

Tests were made also on the distance where the satellite-seeds were formed. In this case, when we fixed one location in the disk where satellites can form (e.g. at $50 \rm{M_{Uranus}}$, at $100 \rm{M_{Uranus}}$, and at $150 \rm{M_{Uranus}}$), we found the number of survived satellites are larger when the seeds were all placed at $150 \rm{M_{Uranus}}$ than at $50 \rm{M_{Uranus}}$, this is just because excluding the outer CPD will of course result in more compact moon-system, that can accomodate less satellites. Because one has no apriori knowledge on where the current satellites were formed, we randomized the satellite-seed positions in our model presented in this paper.

\subsection{Neptune}

As in the case of Uranus, the nearly equal mass Neptune could also form a CPD when $\rm{T_p} < 500$ K (Fig. \ref{fig:hydro}). The CPD in this case was $1.7\times 10^{-3} \,\rm{M_{Neptune}}$. The mass infall rate to the CPD was found to be $7.4\times 10^{-8} \,\rm{M_{Sun}/yr}$. 

\begin{figure*}
\includegraphics[width=18cm]{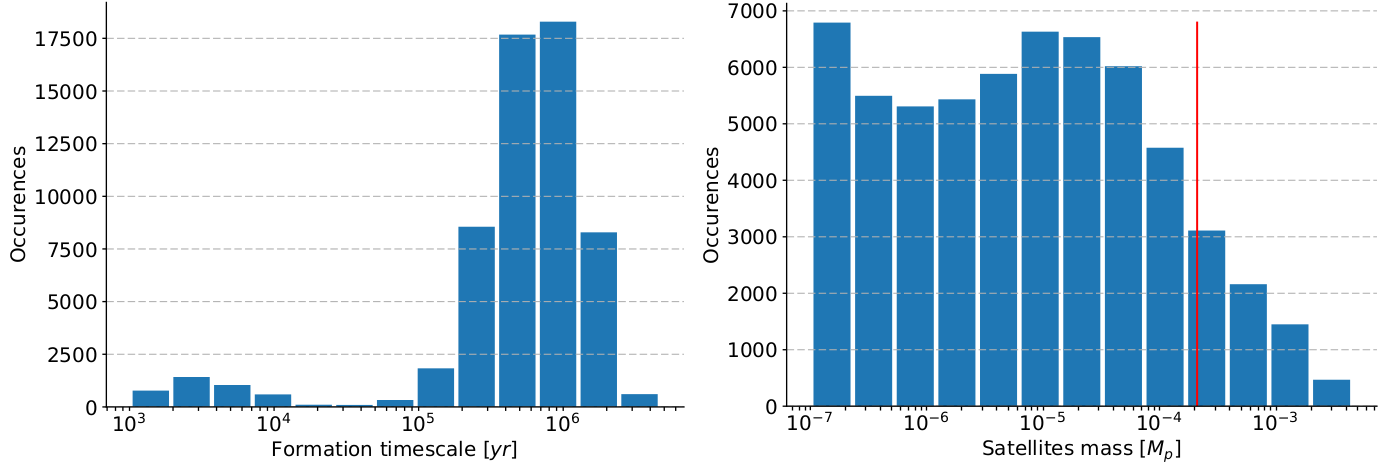}
\caption{Formation timescales of the moons around Neptune in our population synthesis (left). Mass distribution of the formed satellite population around Neptune, where the red line shows the mass of its current one major moon, Triton (right).}
\label{fig:neptune}
\end{figure*}

From the population synthesis, we found similar trends for Neptune as in the case of Uranus. This CPD could also form relatively massive moons in most of the cases. The majority of the population formed over a couple of $10^5$ years (Fig. \ref{fig:neptune} left panel). The entire CPD had a temperature below water freezing point, so this disk also likely forms only icy satellites. The masses of the formed moons were again often in the low mass regime between $10^{-7}$ and $10^{-4} \,\rm{M_{Neptune}}$, i.e. smaller than the mass of Triton (Fig. \ref{fig:neptune} right panel). We remind that Triton is most likely a captured Kuiper-object \citep{McKinnon95}, so it did not form around Neptune, our mass comparison in this case is simply for scaling purposes. However, our result is that a similar moon-system could have formed around Neptune like around Uranus, before the capture of Triton happened. It was already suggested by a dynamical study \citep{Rufu17} that the capture of Triton is only possible, if the original moon-system was similar in mass than the Uranian satellite system.

\section{Conclusion \& Discussion} \label{sec:conclusion}

We investigated CPD- and moon-formation around Uranus and Neptune with combining radiative hydrodynamical simulations with satellite population synthesis. We found that both Uranus and Neptune could form a gaseous disk at the end of their formation, when their surface temperature dropped below 500\,K. These disks are able to form satellites in them within a few hundred-thousand years. The masses of such satellite-systems for both planets were often similar to the current one around Uranus. All the formed moons must be icy in composition, given that they formed in a CPD that has a temperature below water freezing-point. We highlighted that Neptune could have originally a similar satellite system that we see today around Uranus, which was perturbed and got lost by the capture of Triton from the Kuiper-belt, as the dynamical study of \citet{Rufu17} suggested before.

Given that here we showed that the moons of Uranus could have formed in the gaseous CPD around the ice giant, there is no need for a planet-planet impact. Our satellite formation model works for Uranus, no matter whether the axial tilt of the planet was caused by multiple smaller impacts before the satellites formed \citep{Morbidelli12}, or by secular resonance with a larger mass moon \citep{BL10}, that we can even form within our CPD. Unlike the impact scenarios, our gaseous disk naturally forms icy, prograde moons without the need of fine tuning.

Since it is possible to form satellites around ice giants, satellite-formation seems to be more frequent than it was previously thought. This is particularly exciting for exomoon-hunting, because Neptune-mass exoplanets are among the most common mass-category of exoplanets, if they can also form satellites, there must be a much larger population of exomoons than previously considered. Moreover, icy moons in our Solar System are the main targets to search for extraterrestrial life (Europa in case of Jupiter, \citealt{Greenberg11,Sparks17}, and Enceladus in case of Saturn, \citealt{Parkinson08}), hence a larger amount of icy satellites of the similar kind means potentially larger sample of habitable words.

\acknowledgments

We thank for the anonymous referee for their useful suggestions to improve this manuscript. This work was funded by the Swiss National Science Foundation (SNSF) Ambizione grant PZ00P2\_174115. The simulations were done on the ``Moench" cluster hosted at the Swiss National Computational Centre.



\end{document}